%% file: main.tex
\documentclass[a4paper,11pt]{article}
\usepackage{pos}
\usepackage{wrapfig}
\usepackage{lipsum}
\usepackage{notes2bib}

\let\oldcite\cite
\renewcommand{\cite}[1]{\mbox{\oldcite{#1}}}

\newcommand{\fhead}[1]{\vspace{10pt}\goodbreak \noindent \textbf{#1}:\vspace{5pt}\nobreak\par}

\newcommand{\fheadd}[1]{\vspace{5pt}\goodbreak\indent \textbf{#1:}\quad}

\def\traditional{{\tt traditional}}
\def\pnSRC{{\tt pnSRC}}
\def\baseSRC{{\tt baseSRC}}

\def\chidat{\chi^2/N_{\rm data}}

\def\beq{\begin{equation}}
\def\eeq{\end{equation}}
\def\bea{\begin{eqnarray}}
\def\eea{\end{eqnarray}}

\def\collab{\noindent
Performed in collaboration with 
A.W.~Denniston, T.~Je\v{z}o, A.~Kusina, N. Derakhshanian,
P.~Duwent\"aster, O.~Hen, C.~Keppel, M.~Klasen, K.~Kova\v{r}\'{i}k,
J.G.~Morf\'{i}n, K.F.~Muzakka,  E.~Piasetzky, P.~Risse,
R.~Ruiz, I.~Schienbein,  J.Y.~Yu.
\null \vspace{5pt}
}

\def\smu{\affiliation{Department of Physics, Southern Methodist University,
    Dallas, TX 75275-0175, U.S.A.}}

\title{Bridging the Gap:  \\ 
{\Large Connecting Atomic Nuclei  to Their Quantum Foundations}\footnote[2]{\collab}}
\ShortTitle{Bridging the Gap: Connecting Atomic Nuclei  to Their Quantum Foundations}

\author*{{Fredrick Olness}%
}
\smu

\emailAdd{olness@smu.edu}

\abstract{
We extend the QCD Parton Model analysis by employing a factorized nuclear structure model that explicitly accounts for both individual nucleons and correlated nucleon pairs. 
This novel framework establishes a paradigm that directly links the nuclear physics description of matter (in terms of protons and neutrons) to the particle physics schema (in terms of quarks and gluons). 
Our analysis of high-energy data from lepton Deep-Inelastic Scattering, Drell-Yan, and W/Z production simultaneously extracts the universal effective distribution of quarks and gluons inside correlated nucleon pairs, and their nucleus-specific fractions. 
The successful extraction of these universal distributions marks a significant advance in our understanding of nuclear structure, as it directly connects nucleon-level and parton-level quantities.
} %

\FullConference{XXXII International Workshop on Deep Inelastic Scattering and Related Subjects (DIS2025)\\
24-28 March, 2025\\
Cape Town, South Africa\\}

\begin{document}
\maketitle
\nocite{nCTEQ:2023cpo,Kovarik:2015cma,Klasen:2025ekj}

\input{figs}
\figone
\fhead{A Novel Paradigm}

Subatomic systems—including nucleons and atomic nuclei, as well as dense astrophysical matter—derive their fundamental properties from the complex, many-body interactions among their constituent quarks and gluons, all of which are precisely described by the theory of Quantum Chromodynamics (QCD). 

We present a novel framework that combines information from nuclear Parton Distribution Functions (nPDFs) with Short-Range Correlations 
(SRCs)\footnote{%
For a complete list of references, please refer to those cited in  Ref.~\cite{nCTEQ:2023cpo}}.
This approach directly establishes a crucial link between partonic dynamics (quarks and gluons) and nuclear dynamics (protons and neutrons).

The PDFs are a key component that allows us to describe the complex internal structure and interactions of nuclei. Notably, the nuclear PDFs are not a simple sum of the PDFs for free protons and 
neutrons~\cite{Kovarik:2015cma,Klasen:2025ekj}.
Instead, they demonstrate significant nuclear modification effects (such as the EMC effect, shadowing, and antishadowing) that alter the momentum distributions of quarks and gluons inside a nucleus relative to those in individual nucleons.

\fhead{nPDF Framework}

We adopt a model-agnostic approach by focusing on the broad-scale features common to modern, high-resolution nuclear structure models, thereby minimizing dependence on specific model details.
The parameterization framework for the nuclear Parton Distribution Functions (nPDFs) differs significantly from those employed in conventional global analyses. Nevertheless, it achieves a notably good description of the currently available nuclear data, as we shall demonstrate.

We construct our nPDF as a linear combination of a free-nucleon PDF (which accounts for the quasi-free nucleons) and a universal Short-Range Correlation (SRC) PDF that describes the quark and gluon distributions inside an SRC pair:
\begin{align}
 f_i^A(x,Q) =
 \frac{Z}{A} \left[ (1-C_p^A)    f_i^p(x,Q) +C_p^A    f_i^{\mathrm{SRC}\, p}(x,Q) \right] 
  +
  \frac{N}{A} \Big[ (1-C_n^A)    f_i^n(x,Q) + C_n^A    f_i^{\mathrm{SRC}\, n}(x,Q)  \Big]  .
  \nonumber 
\label{eq:4fSRC}
\end{align}
Here, $f_i^A$ is the nPDF of parton type  $i$ (gluon or quark flavors) in a nucleus with mass number {$A$}, carrying momentum fraction $x$ at energy scale $Q$.  $f_i^{p,n}$ and 
$f_i^{\mathrm{SRC}\, {p,n}}$ are the PDFs of the free-nucleon and of the modified nucleon in an SRC pair, respectively. 
We implicitly assume that $f_i^{\mathrm{SRC}\, {p,n}}$ can be defined via a collinear factorization framework, 
and that the proton and neutron distributions are related by 
isospin.\footnote{%
We will assume isospin symmetry to relate the proton and neutron nPDFs. Therefore, in the following discussion, we will simply use $f_i^{p}$ and $f_i^{\mathrm{SRC}\, {p}}$ to identify the free-nucleon and  SRC  nPDFs, respectively.} %
Therefore, we apply the tools from perturbative QCD used for free-nucleon PDFs to arrive at the physical predictions.

A crucial characteristic of SRCs is their universality across the full nuclear spectrum, meaning the short-distance dynamics of the correlated pairs are largely independent of the specific nucleus.

We keep a model-independent approach as to the number of SRC pairs and their isospin structure. 
The nuclear structure dependence is fully encapsulated in the fraction of nucleons in SRC pairs, denoted by $C^A_{p,n}$~~\cite{nCTEQ:2023cpo,Paakkinen:2025pcw}.
In fact, these nuclear structure parameters  $C^A_{p,n}$ will be
independently determined in our nPDF analysis for the first time ({\it cf.},~Fig.\,\ref{fig:one}), and tested for consistency with independent results
from specific nuclear structure studies.
\fhead{nPDF Analysis:}
\figChiTable
The analysis utilized the full set of available  data on nuclear lepton DIS, Drell-Yan processes, and $W$ and $Z$ boson
production.
The corresponding theoretical predictions were obtained at next-to-leading order (NLO) in QCD.
The DIS data primarily constrain the $u$ and $d$ quark and anti-quark distributions. In contrast, the $W$ and $Z$ boson production data from LHC proton-lead collisions also constrain
strange quark and gluon distributions down to lower momentum fractions of $x\,{\sim}\,10^{-3}$.

The energy-scale dependence is accounted for using the DGLAP evolution equation, which also helps constrain the gluon distribution via the $Q$-dependence of DIS data.
The parton number and momentum sum rules are guaranteed to be satisfied separately for the free-nucleon PDFs ($f_i^{p}$) and for the SRC PDFs ($f_i^{\mathrm{SRC}\, p}$).
Because each component already satisfies these fundamental rules, their linear combination (which forms the total nuclear PDF) must also satisfy the sum rules. This holds independently of the specific values used for the proton and neutron SRC fractions \mbox{($C_p^A$ and $C_n^A$).}

The free-nucleon PDFs  $f_i^{p}$ are set to the nCTEQ15 proton distributions~\cite{Kovarik:2015cma}.
We have also explored other proton sets from different groups and obtained comparable results.
The SRC nucleon PDFs  $f_i^{\mathrm{SRC}\, p}$ use the same functional form as  $f_i^{p}$ with 21 shape parameters. 
We perform two independent analyses where 
i)~$C^A_p$ and  $C^A_n$ are allowed to vary freely, 
and ii)~where we assume proton-neutron SRC dominance,
i.e.~$C^A_p=\frac N Z \times C^A_n \equiv C^A$. 
We refer to these fits as the {\tt baseSRC} and {\tt pnSRC} fits, respectively. 
For comparison, we also repeated the traditional (mean-field-like) analysis of Ref.\,\cite{Segarra:2020gtj} using the same dataset, which we refer to as the \traditional\ fit.

\null\vspace{-10pt}
\fhead{Analysis Results}\vspace{-8pt}

\fheadd{The $\chi^2$ Results}
 The resulting fit quality in terms of  $\chi ^2$   are
listed in Table~\ref{tab:fits} for each data type separately, and for all data combined.
The SRC fits result in overall  
$\chi_{{\rm tot}}^{2}/N_{\rm DOF}$ values  
appreciably better than for the \traditional\ fit; 
this fully takes into account the additional SRC ($C_{p,n}^A$) 
parameters.

\fheadd{nPDF Results}
The nPDFs obtained from the SRC approach (not shown) are generally consistent with those from the \traditional\ fits within the estimated uncertainties.
The SRC approach not only yields nPDFs that closely reproduce the \traditional\  results (including the most well-constrained valence distributions) but also achieves an improved $\chi^2/\mathrm{dof}$. This fit successfully reproduces all data across the full range, corresponding to an $x$ range of approximately $10^{-3}$ to $0.75$. 

Given that the initial nPDF parameterization in the SRC framework differs fundamentally from the \traditional\ functional form, achieving results comparable to those of the latter while improving data description is a noteworthy achievement.

\fhead{The  $C_{p,n}^A$ Coefficients}
Figure\,\ref{fig:one} shows the extracted  $C_p^A$ and  $C_n^A$ coefficients as determined by the global {\tt baseSRC}.
The coefficients show logarithmic growth with the nuclear mass number~$A$, starting from ${\sim}5\%$ for He and reaching
${\sim25}\%$  for Pb. 
\fheadd{Figure~\ref{fig:one}-a)}
In the top panel of Fig.~\ref{fig:one}-a)  we display  $\{C_n^A,C_p^A\}$ for the \baseSRC\ fit. 
We also show a rescaled coefficient, $(N/Z)C_n^A$, that accounts for the neutron excess when present. 
The uncertainties of  $\{C_n^A,C_p^A\}$ are obtained by evaluating the $\chi^2$ while scanning across the parameters, and reflect, in part, the number of data points available for each separate nucleus. 
The lines are simple logarithmic fits to show the trend in $A$.

We first observe that the slopes of the proton fraction ($C_p^A$) and the neutron fraction ($C_n^A$) of nucleons in  SRC pairs are roughly similar when compared against the nuclear mass number $A$.
However, after accounting for the neutron excess (by multiplying $C_n^A$ by the factor $N/Z$), we find that the slopes of $C_p^A$ and the rescaled $(N/Z)C_n^A$ become essentially the same within the estimated uncertainties. 
This result indicates that protons and neutrons participate in the SRC interactions in approximately equal numbers, consistent with the hypothesis that SRCs are dominated by proton–neutron pairs.

Furthermore, the observation that $C_p^A$ and the rescaled $(N/Z)C_n^A$ exhibit similar behavior explains why imposing the constraint $C_p^A = (N/Z)C_n^A$ in the \pnSRC\ fit results in only a minimal increase in $\chi^2/\mathrm{dof}$ (from 0.80 to 0.82) relative to the \baseSRC\ fit, where $C_p^A$ and $C_n^A$ are treated independently.

We note that $\{C_n^A,C_p^A\}$ exhibit a scaling freedom on their absolute value due to a freedom in the parameterization \cite{Paakkinen:2025pcw}. Although no evidence of this scaling freedom was identified in our results, we focus the following discussion on the relative size of the coefficients instead of their absolute values.

\fheadd{Example Nuclei}%
For the \baseSRC\ fit, the $\{C_p^A, C_n^A\}$ vary independently without constraints; hence, the observed trends are purely due to the data. 
We previously noted that when the obtained $\{C_p^A, C_n^A\}$ values are adjusted for the neutron excess present in heavy nuclei, the resulting effective number of protons and neutrons in SRC pairs is approximately equal.
We will now illustrate this specific finding for the cases of gold and lead nuclei. 

The estimated uncertainties for $\{C_p^A, C_n^A\}$ are displayed in Fig.~\ref{fig:one} for the case of gold and lead are roughly~$\pm 5\%$.
For gold,  ${}^{197}_{79}\rm Au$,  we obtain $C_p^A{=}0.256$, $C_n^A{=}0.177$. 
This implies $79{\times}C_p^A{\simeq} 20.2 $ protons
and $118{\times} C_n^A{\simeq} 20.9$ neutrons are (effectively) modified by the SRC.
Likewise for the case of lead, ${}^{208}_{82}\rm Pb$, 
we have $C_p^A{=}0.295$, $C_n^A{=}0.202$, which implies 
$82{\times}C_p^A{\simeq} 24.2 $ protons
and $126{\times} C_n^A{\simeq} 25.5$ neutrons are (effectively) modified.
This exemplifies the above observation that once the $C^A$-coefficients
are adjusted to account for the neutron excess, the effective number of protons and neutrons within SRC pairs are approximately equal.

\fheadd{Proton-Neutron SRC Dominance}
More generally, we find the above pattern or proton-neutron pairing is evident throughout 
the range of  Fig.~\ref{fig:one}; we fit the $\{C_p^A, C_n^A\}$ values with a simple 
2-parameter function, $f(A)= a\log(A)+b$, to better display the trends. 
In the top panel of Fig.~\ref{fig:one} we observe $C_p^A$ (blue line) and $C_n^A$ (gray line)
are both increasing with $A$, but with differing slopes. 

To account for the abundance of neutrons for non-isoscalar nuclei, 
we modify the original $C_n^A$ values by a factor of $N/Z$,  
and this yields the orange line with a slope much closer to the $C_p^A$ slope (blue line).
By comparing the rescaled neutron fraction line, $(N/Z)C_n^A$, with the original proton fraction line, $C_p^A$, we gain insight into two key aspects of the SRCs.
\begin{enumerate}

\item Proton-Neutron SRC Dominance: The overlap directly shows the extent to which the effective number of protons and neutrons participating in SRC interactions are comparable, after accounting for the nuclear neutron excess.

\item Nuclear $A$-Dependence: The similarity in their slopes reveals the extent to which the dependence of these parameters on the nuclear mass number ($A$) is comparable, highlighting their universality.

\end{enumerate}
These observations are consistent with the hypothesis that the SRC pairs are dominantly proton-neutron combinations. 

\fheadd{Figure~\ref{fig:one}-b)}
Focusing on the lower panel, Figure\,\ref{fig:one}-b)  
we display the $C_p^A$ coefficients for the \baseSRC.
We also show the logarithmic fits for both \baseSRC\ and \pnSRC\, and they closely coincide. 
Additionally, we overlay predictions from   independent, extractions from
both nuclear structure calculations~\cite{Cruz-Torres:2019fum} and quasi-elastic electron scattering measurements~\cite{CLAS:2005ola,Fomin:2011ng,Schmookler:2019nvf}. 

This behavior is consistent with the observed \mbox{$pn$-dominance} previously established in nuclear structure studies~\cite{Subedi:2008zz,Piasetzky:2006ai}.
Consequently, the agreement between the {\tt baseSRC} and {\tt pnSRC} fits provides an initial indication of consistency between quark–gluon–level analyses and traditional nuclear structure observations.

\fhead{Conclusion}

We have presented a novel paradigm that directly links nuclear structure information to parton (quark and gluon) dynamics. 
This framework enables the first direct extraction of nPDFs that is fully consistent with independent nuclear structure determinations. 

This analysis yields a data description that is similar to or better than that of the traditional parameterization and enables a meaningful physical interpretation of the fit in terms of the nuclear dynamics.
The study successfully determines both the standard ``average'' nPDFs (which can be directly compared with traditional nPDF fits), and the universal distribution of partons in  SRC nucleon pairs, along with the fractions of such SRC pairs. 

The agreement between the SRC fractions derived here and those obtained from low-energy quasielastic measurements establishes a clear connection between high-energy partonic behavior and low-energy nuclear structure, marking a significant step toward understanding nuclei in terms of partonic QCD processes.

The SRC framework provides additional flexibility by allowing one to (i) follow the traditional approach using averaged nucleon distributions, or (ii) construct a more detailed initial-state description in which each nucleon is represented by either a free-nucleon PDF or an SRC-modified PDF, depending on its correlation state.

Furthermore, it is noteworthy that the SRC parameterization (in which the dependence of $A$ and $x$ is factorized) produces an excellent description of the data. The conceptual simplicity of this successful parameterization is striking.

\fhead{Acknowledgment}

This study was performed in collaboration with members of the nCTEQ Collaboration and Or Hen's MIT research group. 
 We are grateful to Tim Hobbs, Jeff Owens and Efrain Segarra
for valuable discussion.
This work was supported by 
the U.S.\ DoE  Grant No.~DE-SC0010129,
and 
by the  U.S.\ DoE, Office of Science, Office of Nuclear Physics, 
within the framework of the Saturated Glue (SURGE) Topical Theory Collaboration.
This work was performed in part at the Aspen Center for Physics, which is supported by National Science Foundation grant PHY-2210452.

\null\vspace{-25pt}

\bibliographystyle{apsrev4-1}
\bibliography{extra,main}

\end{document}

%% file: figs.tex
\def\figone{
\begin{wrapfigure}{R}{0.5\textwidth}
\centering
\null\vspace{-20pt}
\includegraphics[width=0.45\textwidth]{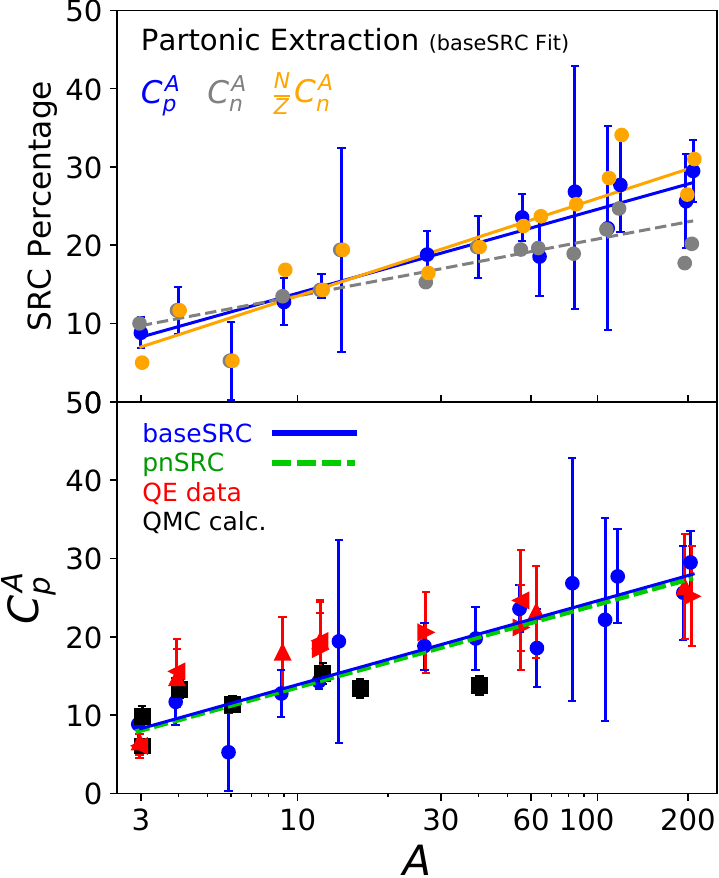}
\caption{(a)~Comparison of nuclear structure parameters $C_p^A$, $C_n^A$, and $(N/Z)C_n^A$ values for the {\tt baseSRC} fit. 
The solid lines represent logarithmic fits to the corresponding quantities.
We show uncertainties only for the $C_p^A$, 
but errors for other quantities are of similar size.
\\
(b)~Comparison of $C_p^A$ values for  the {\tt baseSRC} fit 
and the  SRC abundances extracted from quasi-elastic~(QE)~\cite{CLAS:2005ola,Fomin:2011ng,Schmookler:2019nvf} data
and  Quantum Monte Carlo (QMC)~\cite{Cruz-Torres:2019fum} nuclear calculations. 
The logarithmic fits for  {\tt baseSRC} and {\tt pnSRC} are also shown.
\\[-5pt] \null \hrulefill
\null\vspace{-20pt}
}  %
\label{fig:one}
\end{wrapfigure}
} %
\def\figChiTable{
\begin{wraptable}{R}{0.65\textwidth}
  \null\vspace{-40pt}
  \centering
  \begin{tabular}{|c||c|c|c|c||c||c|}
    \hline   %
    $\chidat$ & DIS  & DY  & $W/Z$  & JLab  & 
    $\chi_{{\rm tot}}^{2}$  
    &
    {\scriptsize \small $\frac{\chi_{{\rm tot}}^{2}}{N_{\rm DOF} }$  }%
    \tabularnewline[4pt]
    \hline 
    \hline    %
    \traditional & 0.85 & 0.97 & 0.88 & 0.72 &  1408 & 0.85
    \tabularnewline
    \hline    %
    \baseSRC   & 0.84 & 0.75 & 1.11 & 0.41 & 1300  &  0.80
    \tabularnewline
    \hline    %
    \pnSRC   & 0.85 & 0.84 & 1.14 & 0.49 & 1350 & 0.82
    \tabularnewline
    \hline 
    \hline    %
    $N_{\rm data}$ & 1136 & 92 & 120 & 336 & 1684 & 
    \tabularnewline
    \hline    %
  \end{tabular}
  \caption{
  The partial $\chidat$ values for the data subsets; 
  the number of data points ($N_{\rm data}$) for each process
  are listed in the bottom row. 
  The \traditional\ fit has 19~shape and 3~$W/Z$ normalization parameters.
  The \baseSRC\ and \pnSRC\ fits have 21~shape, 3~$W/Z$ normalization, and 30~and 19~SRC parameters ($C_{p,n}^A$), respectively.
  In total, there are 1684 data points after cuts. 
  Note the last column ($\chi_{{\rm tot}}^2/N_{\rm DOF}$) fully takes into account the number of fit parameters.
    \\[-5pt] \null \hrulefill 
    \null\vspace{-10pt}
  }
  \label{tab:fits}
\end{wraptable}
} %